# Image Augmentation using a Task Guided Generative Adversarial Network for Age Estimation on Brain MRI

Ruizhe Li[1], Matteo Bastiani[2], Dorothee Auer[2], Christian Wagner[1], and Xin Chen[1]

[1]IMA/LUCID Group, School of Computer Science, University of Nottingham, UK
[2]School of Medicine, University of Nottingham, UK

**Abstract.** Brain age estimation based on magnetic resonance imaging (MRI) is an active research area in early diagnosis of some neurodegenerative diseases (e.g. Alzheimer, Parkinson, Huntington, etc.) for elderly people or brain under-development for the young group. Deep learning methods have achieved the state-of-art performance in many medical image analysis tasks, including brain age estimation. However, the performance and generalisability of the deep learning model are highly dependent on the quantity and quality of the training data set. Both collecting and annotating brain MRI data are extremely time-consuming. In this paper, to overcome the data scarcity problem, we propose a generative adversarial network (GAN) based image synthesis method. Different from the existing GAN-based methods, we integrate a task-guided branch (a regression model for age estimation) to the end of the generator in GAN. By adding a task-guided loss to the conventional GAN loss, the learned low-dimensional latent space and the synthesised images are more task-specific. It helps to boost the performance of the down-stream task by combining the synthesised images and real images for model training. The proposed method was evaluated on a public brain MRI data set for age estimation. Our proposed method outperformed (statistically significant) a deep convolutional neural network based regression model and the GAN-based image synthesis method without the task-guided branch. More importantly, it enables the identification of age-related brain regions in the image space. The code is available on GitHub[1].

## 1  Introduction

Data-driven deep learning methods have achieved the state-of-the-art performance in various medical image analysis tasks, which normally require a large amount of data for model training. However, collecting a large quantity of data with high-quality annotation is a big challenge in medical imaging. Data augmentation is one of the widely used methods to address this problem. Classical image augmentation techniques (e.g. geometric transformation, colour space augmentation, random erasing, etc.) are not effective enough to simulate different imaging variations realistically. Therefore, more sophisticated methods based on generative modelling were proposed by researchers [1][2][3].

---

[1] https://github.com/ruizhe-l/tgb-gan

The basic idea of generative modelling based image augmentation is to learn the underlying data distribution from a set of real images, then new images can be synthesised by sampling the learned distribution. In 2014, generative adversarial network (GAN), proposed by Goodfellow et al. [4], was utilised for generating high-quality images from a random latent vector. Subsequently, many image reconstruction and image-to-image translation methods were proposed based on GAN. In 2016, Larsen et al. proposed VAE-GAN which combined variational autoencoder (VAE) [5] and GAN by integrating the decoder of VAE and the generator of GAN. VAE-GAN learns to minimise the dissimilarity of the latent features of the reconstructed image and the input image, as well as a classical GAN-based adversarial loss. Similarly, another image to image translation method was proposed by Isola et al. in 2017 [6]. The authors replaced the generator in GAN with an encoder-decoder based deep convolutional neural network (CNN). A pixel-wise loss was also added to measure the similarity of the original images and the reconstructed images. Both methods have achieved superior performance in generating realistic images. However, this type of methods can only synthesise images from the same distribution of the input dataset without the capability of specifying certain characteristics in the image (e.g. synthesise face images for the same person but at a different age).

To overcome this problem, several methods were proposed in recent years. In 2017, a conditional generative adversarial network (CGAN) [7] based face generation model was proposed by Antipov et al. [8]. It firstly trains an encoder on a dataset that is generated from a pre-trained age-condition GAN. Then a face recognition neural network is used to minimise the distance between real images and synthesised images to preserve the person's identity. Subsequently, an image of certain age of the same person can be generated by conditioning on a given age. Another face ageing method was proposed by Wang et al. in 2018 [9]. It also contains an identity-preserved control loss which minimises the distance between the feature maps of the real and synthesised images that were obtained from a pre-trained feature extraction model. Additionally, a pre-trained age classification model is used to generate an age classification loss for forcing the generated face being in the specified age group.

In medical imaging applications, Xia et. al [10] proposed to learn a joint distribution of brain images and ages using GAN, and also added an identity-preserve loss to generate more age-related images. Besides, several image augmentation methods proposed recently that applied feature interpolation in latent feature space to generate specific target images [11][12]. However, most of these studies focus on improving the quality of the synthesised images and rarely simultaneously investigate the effects of using the synthesised images for down-stream tasks, such as classification of a disease. Arguably, a globally realistically-looking synthesised image does not necessarily contain useful image features for classifying a disease, as most of the information are in a local image region and are concealed by superposed and more distinguishable features (e.g. age and gender variations).

In this paper, we propose a GAN based image synthesis method which aims to generate images that can be used to improve the performance of an age estimation task. The main contributions of this paper are summarised as follows. (1) We integrate a task guided branch to the end of the generator in GAN, and a task-specific loss is added to the conventional GAN loss for guiding the learned low-dimensional latent space to be more task-specific. Hence the synthesised images are also more task-specific. (2) We evaluated our method on a public brain MRI dataset for the task of age estimation. The task-guided branch in this case is an age regression model. We have demonstrated the effectiveness of the proposed method by ablation studies. The results have shown that our method is able to generate age-specific MRI images and therefore boosting the performance of the age estimation model.

## 2 Methodology

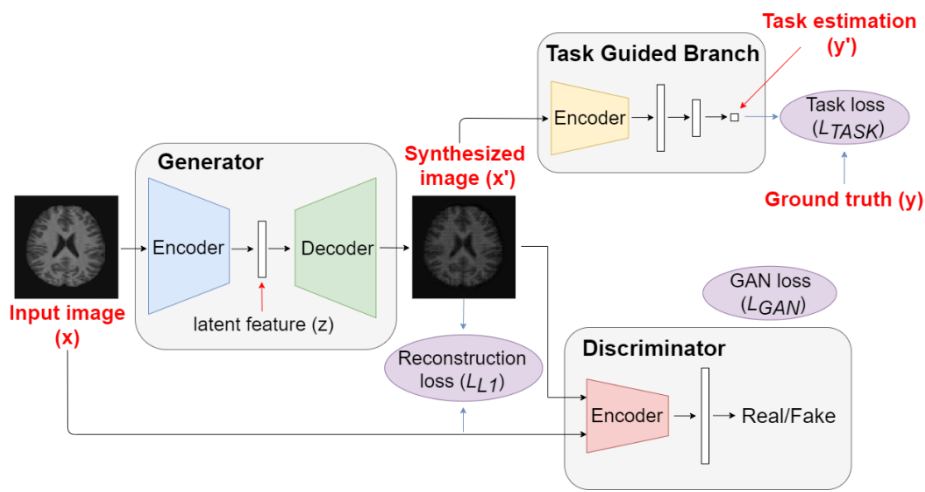

**Fig. 1.** Overview of the proposed method.

The overview of our method is depicted in Fig. 1. The proposed method mainly consists of three network blocks: a generator, a discriminator and a task-guided branch. The generator and discriminator form an image reconstruction network, which is similar to the GAN architecture used in [6]. The task-guided branch is concatenated to the end of the generator, which is a regression network for age estimation. The whole network is trained by optimising an objective function that consists of three terms: a pixel-wise reconstruction loss $L_{L1}$, a GAN loss $L_{GAN}$ and a task-guided loss $L_{TASK}$. The details of each components are described in the following subsections.

### 2.1 GAN based Image Synthesis

To achieve image synthesis, we use a modified 3D image to image GAN [6]. To overcome the mode collapse problem in conventional GAN, we use the loss proposed in WGAN-GP [13]. The generator of our GAN model consists of an encoder and a decoder. Both of them have 5 layers of feature maps by applying one 3×3×3 convolutional operations and one rectified linear unit (ReLU) at each layer. In the encoding path, a max-pooling with a stride of 2 is applied between two layers for feature map downsampling. The number of feature channels for the 5 encoding layers are 64, 128, 256, 512 and 512 respectively.

Correspondingly, a de-convolutional operation is performed in the decoder between two consecutive layers to up-sample the feature maps. The number of feature channels for the 5 decoding layers are 512, 512, 256, 128 and 64 respectively. A 1×1×1 convolutional layer is applied at the end of the decoder to convert the dimensions of the output to the same as the input. The discriminator has the same structure as the generator's encoder, but a 1×1×1 convolutional layer is appended to the end for reducing the dimensionality for image discrimination. The cost function of our image synthesis network consists of an adversarial loss $L_{GAN}$ (Eqn. (1)) that has been used in WGAN-GP [13], and a pixel-wise reconstruction loss $L_{L1}$ (Eqn. (2)).

$$L_{GAN}(G, D) = \mathbb{E}_x[D(G(x)) - \mathbb{E}_x[D(x)]] \tag{1}$$

$$L_{L1}(G) = \mathbb{E}_x[|x - G(x)|] \tag{2}$$

where $G: x \rightarrow \hat{x}$ is the generator that maps an input image $x$ to its target reconstruction image $\hat{x}$. $D$ indicates the discriminator that classifies if an image is real or fake. The generator intends to fool the discriminator by producing realistic images, and the discriminator aims to identify the fake ones from the real images. The reconstruction loss $L_{L1}$ minimises the pixel-wise intensity differences between the real and the synthesised images. A weight $\alpha$ is applied to balance the $L_{L1}$ and $L_{GAN}$ losses (Eqn. (3)). The image synthesis network is therefore trained by optimising the following objective:

$$\min_G \max_D L_{GAN}(G, D) + \alpha L_{L1}(G) \tag{3}$$

### 2.2 Task-guided Branch

For the age estimation application, we intend to synthesise MRI images of a particular age if not enough images are available for that age group. The GAN-based image synthesis network introduced in section 2.1 can be used to generate new images by sampling the learned latent feature space (output of the encoder in the generator) and then inputting into the decoder in the generator. Ideally, if an age-specific manifold in the latent space can be identified, an MRI image of a particular age can be generated by linear interpolation in the manifold using the latent feature vectors of images in nearby age groups. However, the latent space is high dimensional, which controls variety of image characteristics (e.g. geometric, intensity, context, gender etc.). It makes it extremely difficult to identify the age-specific manifold.

Hence, we propose to integrate an age-specific branch to the end of the generator in the GAN model that guides the latent feature learning to be more age-specific. In our age estimation application, as a widely used method in image analysis [10], a 3D VGG-like [14] regression network is added to the GAN. It consists of 5 down-sampling layers and 4 fully connected (FC) layers. Each down-sampling layer includes two 3×3×3 convolutional operations, each of them is followed by a ReLU activation function. As a commonly used improvement, a residual block [15] is used in each layer for faster convergence. The number of feature channels in the 5 down-sampling layers are 16, 32, 64, 128 and 256 respectively. The feature map of the last down-sampling layer is flattened to a vector and fed into the FC layers. Each FC layer is also followed by a ReLU function, and the number of units in the FC layers are 2048, 1024, 64, 1, respectively. The final FC layer produces the estimated age. The *L2* loss is used to measure the error between the estimated age and the ground truth age, as expressed in Eqn. (4).

$$L_{TASK}(G) = \mathbb{E}_{G(x)}[(y - R(G(x)))^2] \tag{4}$$

$y$ is the ground truth age for an input image $x$. $R: G(x) \rightarrow \hat{y}$ is the estimated age using the regression model $R$. $G(x)$ is the synthesised image from the previously described GAN model. Both the GAN model and the regression branch are optimised simultaneously end-to-end. By combining both the GAN loss and the regression loss (with a weight $\beta$), the final objective function of our method is:

$$\min_G \max_D L_{GAN}(G, D) + \alpha L_{L1}(G) + \beta L_{TASK}(G) \tag{5}$$

### 2.3 Latent Space Interpolation for Image Synthesis

After model training, we synthesise new images by interpolating the latent feature space. Specifically, to generate an image of a particular age $y^s$, we randomly select two images (one younger and one older) in the training set with an age difference of less than 1 to $y^s$. The synthesised image $S_{img}$ is generated using Eqn. (6) and the associated age is calculated using Eqn. (7).

$$S_{img} = G_{decoder}(\varepsilon G_{encoder}(x^i) + (1 - \varepsilon) G_{encoder}(x^j)) \tag{6}$$

$$y^s = \varepsilon y^i + (1 - \varepsilon) y^j \tag{7}$$

where $x^i$ and $x^j$ are the two randomly selected images with the age of $y^i$ and $y^j$ respectively. $\varepsilon$ is a randomly determined value between 0 and 1. $G_{encoder}$ is the trained encoder in the generator that generates the latent feature for a given image, and $G_{decoder}$ is the decoder in the generator that generates a synthesised image from the interpolated latent feature vector. Many pairs of $S_{img}$ and $y^s$ can then be generated and used as the additional training data for age estimation model training.

## 3    Experiments and Results

### 3.1    Dataset

We evaluated the proposed method using a public brain MRI dataset from the Autism Brain Imaging Data Exchange (ABIDE) I [16] and II [17]. Standard minimal pre-processing was run on the data, following the steps described in [18], including brain extraction, bias field correction and nonlinear registration [19] [20] [21] to the MNI template space. The dataset used in this paper contains 1150 images of healthy subjects with age from 6 to 24. Note that, after the pre-processing steps, variations of the brain volume and geometric differences across subjects were removed. This prevents the model from learning the most distinguishable feature (i.e. brain volume) for age estimation. Features like brain volume can be easily obtained by image segmentation and integrated to an age estimation model. Here we train the model to learn local features that is much more difficult to be observed by human. After these pre-processing steps, the image size is 182×218×182 voxels, and we resized them to 96×96×96 voxels due to the limitation of GPU memory. The main aim of the experiments was to demonstrate the effectiveness of the proposed task-guided branch rather than to achieve the state-of-the art age estimation performance, hence the down-sampled images were used to reduce computational cost and memory consumption. We also applied the zero-mean and min-max methods as pre-processing steps to normalise the image intensity to the range of [0, 1]. To avoid model bias, we balanced the numbers of male and female for each age group (age interval of 1 year), resulting in 862 images (431 male and 431 female). Note that the number of images for each age group is not balanced (e.g. 6 samples for age 5 to 6, 86 samples for age 9 to 10). Subsequently, we divided them into training, validation and testing sets with the ratio of approximately 75%, 5% and 20% (i.e. 640, 62 and 160) respectively. For a fair comparison, all experiments performed in this paper used the same validation and test sets.

### 3.2    Experiments

For method comparison, we firstly trained a baseline regression model (named as REG) using the VGG-based architecture as described in Section 2.2 for age estimation based on the 640 training images.

As the second comparative method, we trained a GAN-based model (Section 2.1) without the age regression branch. As the number of images in different age groups are highly unbalanced, we generated synthesised images using the method described in Section 2.3 to ensure at least 50 images for each age group after combining the real and synthesised images, resulting in a total of 940 images for training. Subsequently, an age regression network (same as REG) was trained using the new dataset (940 images). We name this method REG-GAN.

For our proposed method, we trained the GAN-based model with a regression branch by optimising the objective function in Eqn. (5). Synthesised images were then generated for each age group in the same way as the REG-GAN method, and followed by training an age estimation model using the new dataset. We name our proposed method REG-GAN-TGB.

For a fair comparison, both REG-GAN and REG-GAN-TGB were trained for 100 epochs with the batch size of 20. Adam optimizer [22] was used. The learning rate was 0.0001 and multiplied by 0.8 for every 10 epochs. The parameter tuning was performed using the validation set. As a starting point, we tried to make each part of the loss function (Eqn. (5)) having similar values (around 0.5) after several epochs. Several experiments were performed by varying the weights around that initial values using the validation set. The final values of $\alpha$ and $\beta$ in Eqn. (5) were experimentally determined as 10 and 0.1 respectively.

All of the final age estimation models in REG, REG-GAN and REG-GAN-TGB were trained for 200 epochs with the batch size of 20. The learning rate was 0.0001 and decayed by multiplying 0.8 in every 10 epochs.

### 3.3    Results

We report quantitative results measured by Mean Squared Error (MSE) and Mean Absolute Error (MAE) of age estimation, and the corresponding image reconstruction loss ($L_{L1}$) of GAN, as listed in Table 1. The age estimation errors of 160 testing images using the three methods for each of the age groups are presented in Fig. 2. The error bars indicate the mean and standard deviation of MSE. The number of real images for training and testing are also presented as bins in Fig. 2.

Table 1 shows that the age estimation result (MSE and MAE) of REG-GAN method is even worse than the baseline REG method. This indicates that a synthesised image using REG-GAN method is not correlated to the associated age, which has a negative impact to the regression model training. In contrast, our proposed method REG-GAN-TGB achieved better performance than the other methods with statistical significance (measured by Wilcoxon signed rank test with p<0.01). It demonstrates the effectiveness of the added task-guided branch, which helps the GAN model to synthesis more meaningful images for different age groups.

**Table 1.** Comparison of our proposed method (REG-GAN-TGB) to the baseline method (REG) and REG-GAN method. Mean $\pm$ standard deviation of $L_{L1}$ loss (image reconstruction error), MSE and MAE of age estimation are reported.

| Method | $L_{L1}$ loss | MSE | MAE |
|---|---|---|---|
| REG | N/A | 5.1922±9.0424 | 1.7060±1.5153 |
| REG-GAN | 0.0168±0.0044 | 6.7092±11.6758 | 1.9356±1.7266 |
| REG-GAN-TGB | 0.0264±0.0066 | **4.0699**±6.2613 | **1.5415**±1.3055 |

From the MSE of age estimation for each of the age groups presented in Fig. 2, it can be seen that the error increases when the number of real images decreases for all three methods. Overall, the proposed REG-GAN-TGB method consistently improved the baseline REG method for most age groups, while the REG-GAN method performed worse than the baseline method for most age groups. The larger errors for some age groups (e.g. 18-22) were due to the limited number of real images for training and image synthesis. It also led to the relatively larger standard deviation values of MSE in Table 1 and Fig. 2.

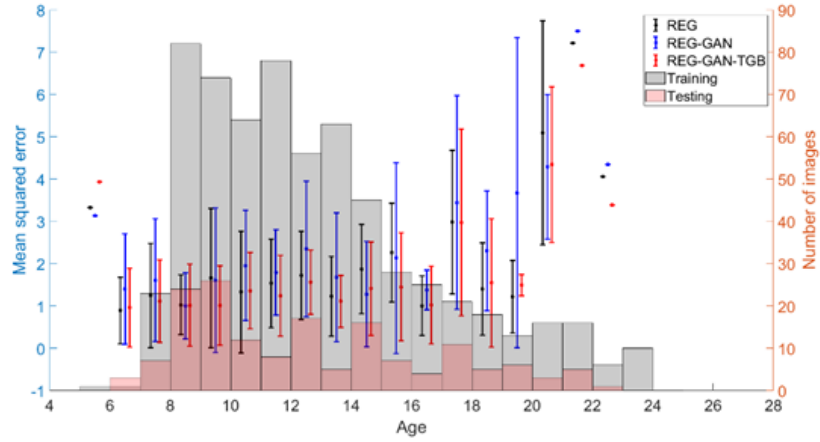

**Fig. 2.** MSE of age estimation for each of the age groups using REG (black), REG-GAN (blue) and REG-GAN-TGB (red). Error bars indicate the mean and standard deviation values. Bins show the number of training (grey) and testing (pink) images of real data.

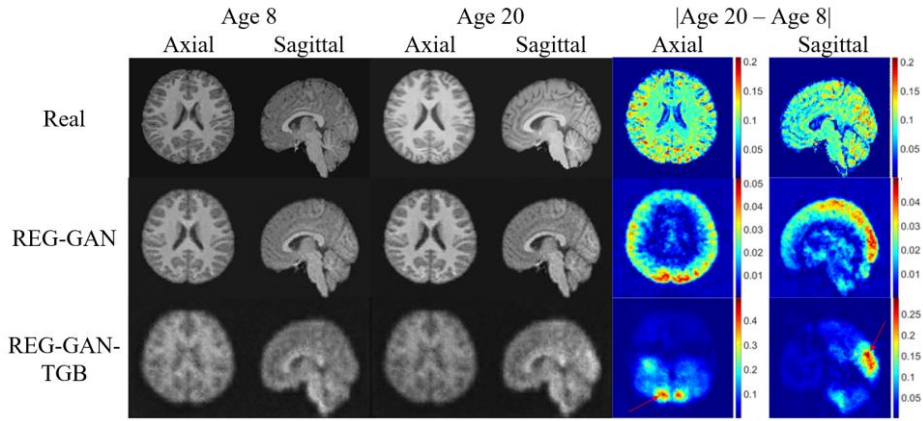

**Fig. 3.** Visual examples of real and synthesised images using REG-GAN and REG-GAN-TGB for age 8 and 20 respectively. |Age 20-Age 8| column shows the absolute intensity difference of the average image of 20 real/reconstructed images between age 8 and 20 for each method.

In Fig. 3 columns 1 to 4, we show some visual examples of real images and synthesised images using the REG-GAN and REG-GAN-TGB methods for age 8 and 20 in both axial view (slice 50) and sagittal view (slice 50). For each of the methods, we also show the image intensity difference of using the average image of 20 real/reconstructed images for age 20 subtracted by the average image of 20 real/reconstructed images for age 8 (column 5 and 6 in Fig. 3). It can be seen that the intensity difference image obtained from the real MRIs does not highlight specific brain regions that reflect the age difference. It is also observed that the visual image quality generated by the REG-GAN method is much better than our method. However, the intensity difference between age 8 and 20 generated by REG-GAN is quite small (maximum of 0.05 out of 1) and mainly highlighted the whole brain border regions. In contrast, although our proposed method generated more blurry images than REG-GAN, it produced a much larger intensity difference (maximum of 0.49 out of 1) in the occipital lobe region than other brain regions (red arrow in Fig. 3), indicating an active change from age 8 to 20. The occipital lobe contains the primary and associated visual cortex which is primarily responsible for visual processing. This finding needs to be further confirmed in brain science field. Overall, we have demonstrated that our image synthesis method not only helps to boost the performance of downstream age estimation task but also extremely powerful to identify age-related image regions. The image synthesis process using our method can be considered as a low pass filtering that effectively extracts the key age-related information.

## 4    Discussion and Conclusions

In this paper, we have proposed an image synthesis solution using a GAN-based model. Based on an age estimation task using 3D brain MRI, we have demonstrated that by integrating a regression branch with an additional loss term to the GAN model, the learned latent feature space and the synthesised images are more age-specific. With the augmented data from our proposed method, the performance of the age estimation model is significantly improved. However, like most machine learning generative models, the synthesised images are still in the same data distribution of the training set, which cannot be used to generate out-of-distribution samples. The main benefit of the proposed solution is to balance the dataset to minimised the bias in model learning.

It is also noteworthy to emphasise that we removed the brain volume variations for the age estimation modelling as we were keen to explore new local image features that are more difficult to be identified by human. Using our method, we have identified the occipital lobe brain region that is highly correlated to the age difference between 8 and 20.

For the comparison of our method to other state-of-the-art methods, we investigated other studies that use brain MRIs for age estimation [23]. Most of them were evaluated on an age range of 18 to 90 years, and the MAE were around 4 to 7. In 2020, He et al. [24] proposed a method to estimate the age for children on a subset of a public brain MRI dataset. The range of age they used was 0-22 years. As an indirect comparison, the MAE of their 3D method on that dataset in the age groups of 6-10, 11-15, 16-22 were 1.12, 1.19, 1.85, respectively. The MAE of our proposed REG-GAN-TGB method on our dataset in the same age range were 1.32, 1.37 and 2.11 respectively.

For future work, we plan to test our method on more datasets and comprehensively compare the performance with other state-of-the-art methods. We will further investigate the impact of the task-guided branch in the latent space and attempt to detect disease biomarkers by visualising different features of the generated images. For example, use this approach in a Parkinson dataset to explore whether the method could again highlight meaningful brain regions that correlate to the disease progression.

# References

1. Shin, H.C., Tenenholtz, N.A., Rogers, J.K., Schwarz, C.G., Senjem, M.L., Gunter, J.L., Andriole, K.P., Michalski, M.: Medical image synthesis for data augmentation and anonymization using generative adversarial networks. In: Lecture Notes in Computer Science (including subseries Lecture Notes in Artificial Intelligence and Lecture Notes in Bioinformatics) (2018). https://doi.org/10.1007/978-3-030-00536-8_1.
2. Frid-Adar, M., Diamant, I., Klang, E., Amitai, M., Goldberger, J., Greenspan, H.: GAN-based synthetic medical image augmentation for increased CNN performance in liver lesion classification. Neurocomputing. (2018). https://doi.org/10.1016/j.neucom.2018.09.013.
3. Pesteie, M., Abolmaesumi, P., Rohling, R.N.: Adaptive Augmentation of Medical Data Using Independently Conditional Variational Auto-Encoders. IEEE Trans. Med. Imaging. (2019). https://doi.org/10.1109/TMI.2019.2914656.
4. Goodfellow, I., Pouget-Abadie, J., Mirza, M., Xu, B., Warde-Farley, D., Ozair, S., Courville, A., Bengio, Y.: Generative adversarial nets. In: Advances in neural information processing systems. pp. 2672–2680 (2014).
5. Kingma, D.P., Welling, M.: Auto-encoding variational bayes. In: 2nd International Conference on Learning Representations, ICLR 2014 - Conference Track Proceedings (2014).
6. Isola, P., Zhu, J.Y., Zhou, T., Efros, A.A.: Image-to-image translation with conditional adversarial networks. In: Proceedings - 30th IEEE Conference on Computer Vision and Pattern Recognition, CVPR 2017 (2017). https://doi.org/10.1109/CVPR.2017.632.
7. Mirza, M., Osindero, S.: Conditional generative adversarial nets. arXiv Prepr. arXiv1411.1784. (2014).
8. Antipov, G., Baccouche, M., Dugelay, J.-L.: Face aging with conditional generative adversarial networks. In: 2017 IEEE international conference on image processing (ICIP). pp. 2089–2093 (2017).
9. Wang, Z., Tang, X., Luo, W., Gao, S.: Face Aging with Identity-Preserved Conditional Generative Adversarial Networks. In: Proceedings of the IEEE Computer Society Conference on Computer Vision and Pattern Recognition (2018). https://doi.org/10.1109/CVPR.2018.00828.
10. Xia, T., Chartsias, A., Tsaftaris, S.A., Initiative, A.D.N., others: Consistent brain ageing synthesis. In: International Conference on Medical Image Computing and Computer-Assisted Intervention. pp. 750–758 (2019).
11. Liu, X., Zou, Y., Kong, L., Diao, Z., Yan, J., Wang, J., Li, S., Jia, P., You, J.: Data augmentation via latent space interpolation for image classification. In: 2018 24th International Conference on Pattern Recognition (ICPR). pp. 728–733 (2018).
12. DeVries, T., Taylor, G.W.: Dataset augmentation in feature space. In: 5th International Conference on Learning Representations, ICLR 2017 - Workshop Track Proceedings (2019).
13. Gulrajani, I., Ahmed, F., Arjovsky, M., Dumoulin, V., Courville, A.: Improved training of wasserstein GANs. In: Advances in Neural Information Processing Systems (2017).

14. Simonyan, K., Zisserman, A.: Very deep convolutional networks for large-scale image recognition. arXiv Prepr. arXiv1409.1556. (2014).
15. He, K., Zhang, X., Ren, S., Sun, J.: Deep residual learning for image recognition. In: Proceedings of the IEEE conference on computer vision and pattern recognition. pp. 770–778 (2016).
16. Di Martino, A., Yan, C.-G., Li, Q., Denio, E., Castellanos, F.X., Alaerts, K., Anderson, J.S., Assaf, M., Bookheimer, S.Y., Dapretto, M., others: The autism brain imaging data exchange: towards a large-scale evaluation of the intrinsic brain architecture in autism. Mol. Psychiatry. 19, 659–667 (2014).
17. Di Martino, A., O'connor, D., Chen, B., Alaerts, K., Anderson, J.S., Assaf, M., Balsters, J.H., Baxter, L., Beggiato, A., Bernaerts, S., others: Enhancing studies of the connectome in autism using the autism brain imaging data exchange II. Sci. data. 4, 1–15 (2017).
18. Alfaro-Almagro, F., Jenkinson, M., Bangerter, N.K., Andersson, J.L.R., Griffanti, L., Douaud, G., Sotiropoulos, S.N., Jbabdi, S., Hernandez-Fernandez, M., Vallee, E., others: Image processing and Quality Control for the first 10,000 brain imaging datasets from UK Biobank. Neuroimage. 166, 400–424 (2018).
19. Jenkinson, M., Bannister, P., Brady, M., Smith, S.: Improved optimization for the robust and accurate linear registration and motion correction of brain images. Neuroimage. 17, 825–841 (2002).
20. Andersson, J.L., Jenkinson, M., Smith, S.: Non-linear optimisation. FMRIB Analysis Group Technical Reports. TR07JA1, (2007).
21. Jenkinson, M., Beckmann, C.F., Behrens, T.E.J., Woolrich, M.W., Smith, S.M.: Fsl. Neuroimage. 62, 782–790 (2012).
22. Kingma, D.P., Ba, J.: Adam: A method for stochastic optimization. arXiv Prepr. arXiv1412.6980. (2014).
23. Sajedi, H., Pardakhti, N.: Age prediction based on brain MRI image: a survey. J. Med. Syst. 43, 279 (2019).
24. He, S., Gollub, R.L., Murphy, S.N., Perez, J.D., Prabhu, S., Pienaar, R., Robertson, R.L., Grant, P.E., Ou, Y.: Brain Age Estimation Using LSTM on Children's Brain MRI. In: 2020 IEEE 17th International Symposium on Biomedical Imaging (ISBI). pp. 1–4 (2020).